\documentclass[12pt]{iopart}
\usepackage{graphicx}
\usepackage{epstopdf,color,cite}
\usepackage{url}

 \definecolor{Black}{named}{Black}
 \definecolor{Blue}{named}{Blue}
 \definecolor{Red}{named}{Red}

\def\la{\mathrel{\mathpalette\fun <}}
\def\ga{\mathrel{\mathpalette\fun >}}
\def\fun#1#2{\lower3.6pt\vbox{\baselineskip0pt\lineskip.9pt
  \ialign{$\mathsurround=0pt#1\hfil##\hfil$\crcr#2\crcr\sim\crcr}}}

\begin{document}

\title[Ultrahigh Energy Nuclei in the Galactic Magnetic Field]{Ultrahigh Energy Nuclei in the Galactic Magnetic Field}

\author{G.~Giacinti$^{1}$, M.~Kachelrie\ss$^{2}$, D.~V.~Semikoz$^{1,3}$, G.~Sigl$^{4}$}

\address{$^1$ {\it AstroParticle and Cosmology (APC), 10, rue Alice Domon et L\'eonie Duquet, 75205 Paris Cedex 13, France}}
\address{$^2$ {\it Institutt for fysikk, NTNU, Trondheim, Norway}}
\address{$^3$ {\it Institute for Nuclear Research of the Russian Academy of Sciences, 60th October Anniversary prospect 7a, Moscow 117312, Russia}}
\address{$^4$ {\it II. Institut f\"ur Theoretische Physik, Universit\"at Hamburg, Luruper Chaussee 149, D-22761 Hamburg, Germany}}

\begin{abstract}
Observations 
are consistent with a significant fraction of heavy nuclei in the cosmic ray flux above a few times 
$10^{19}\,$eV. Such nuclei can be deflected considerably in the Galactic magnetic field,
with important implications
for the search of their sources. We perform detailed simulations of heavy nuclei propagation
within recent Galactic magnetic field models. While such models are not yet sufficiently
constrained to predict deflection maps in detail, we find general features of the distribution
of (de-) magnified flux from sources.
Since in most theoretical models sources of heavy nuclei are located in 
the local large scale structure of galaxies, we show examples of images of several 
nearby galaxy clusters and of the supergalactic plane.
Such general features may be useful to develop efficient methods for source reconstruction
from observed ultrahigh energy cosmic ray arrival directions.
\end{abstract}


\pacs{98.70.Sa, 98.35.Eg}



\section{Introduction}

Recently, the Pierre Auger Observatory presented measurements of the ultrahigh energy cosmic 
ray (UHECR) composition at the highest energies using the width of the $X_{\max}$ distribution of 
air showers. The width of this distribution has relatively small theoretical uncertainties and the 
Pierre Auger data show a clear shift towards a heavier composition at the highest energies, above a few times 
$10^{19}$\,eV~\cite{Collaboration:2010yv}. The analysis of the Yakutsk EAS Array muon data also revealed a 
significant heavy element fraction at energies above $10^{19}$\,eV~\cite{Glushkov:2007gd}.
Prompted by the possibility of a significant contribution of nuclei in the 
UHECR flux, we investigate in the present work the propagation of heavy nuclei in the Galactic 
magnetic field (GMF). Differences between the propagation of protons and 
heavy nuclei in the GMF may provide additional information about
the charge composition of UHECRs. Additional evidence may be particularly valuable, since
the experimental data from composition studies are still controversial: In contrast to the Pierre
Auger and Yakutsk data, the measurements of the HiRes experiment are consistent with a 
proton composition~\cite{BelzICRC}, a result that is in line with preliminary studies of 
the Telescope Array~\cite{TA}.

Most of the existing papers studying deflections of UHECRs in the GMF consider protons or light 
nuclei. Important exceptions are the studies of Refs.~\cite{Harari:1999it,Harari:2000he,Harari:2000az} that were performed for fixed rigidities $E/Z$ for values as low as $E/Z=1$\,EeV. In the same way, Ref.~\cite{Yoshiguchi:2004kd} presented numerical simulations for the propagation of protons with energies $E\geq1$\,EeV. More recently, Refs.~\cite{Takami:2009qz,Vorobiov:2009km} discussed the effect of varying the UHECR 
composition on the correlation of Auger events with active galactic nuclei, while
the effect of magnetic lensing in a particular example of lens geometry was studied in 
Ref.~\cite{Battaner:2010bd}. 

A shift towards a heavier composition would strongly affect the expected UHECR sky distribution. It would imply, even at the highest energies, the existence of strongly magnified and demagnified regions. Moreover, a fraction of the sky would be invisible, depending on the number of observed CRs. The most critical consequences concern the search for astrophysical sources.

The GMF displays a large scale and a random small scale structure. The first one is known as the regular component, for which several analytical models exist, and the second one is the turbulent component. One of the very first models of the regular GMF was proposed by T.~Stanev in Ref.~\cite{Stanev:1996qj}. It describes analytically the GMF structure in the Milky Way disk in terms of  logarithmic spirals, following its spiral arms visible in stars and gas. Other variants of modeling the disk field were suggested by D.~Harari~et~al.~\cite{Harari:1999it}, and by P.~Tinyakov and I.~Tkachev~\cite{Tinyakov:2001ir}. Later, M.~Prouza and R.~Smida built a model which also contains a halo contribution consisting of toroidal and poloidal fields~\cite{Kachelriess:2005qm,PS} (hereafter named the ``PS model''). Some of the latest information on the GMF were taken into account, among others, in References~\cite{Han:2006ci,Han:2009jg,Han:2009ts,Hinshaw:2006ia,Sun:2007mx,Jansson:2009ip,Jiang:2010yc}.
However, as shown in Refs.~\cite{Jansson:2009ip,Waelkens:2008gp}, currently no GMF model can fit in a satisfactory way all experimental data.

At the energies we consider in this work, $E\ga60$\,EeV, and with the usual assumptions on the turbulent GMF~\cite{Harari:2002dy,Tinyakov:2004pw}, we are still far from the diffusion regime even for iron nuclei. In the ballistic regime, the deflections induced by the turbulent GMF are smaller than or at most comparable to deflections in the regular GMF in most parts of the sky~\cite{Tinyakov:2004pw}. For that reason, we restrict the present study to the regular GMF contribution only.

We backtrace here iron nuclei in different regular GMF models and show that sources located in certain parts of the sky do not contribute to the UHECR flux observed at Earth. Magnetic lensing results in magnification and demagnification of the fluxes from individual sources. We quantitatively investigate these effects which can be important in the heavy nuclei UHECR scenario.

We also study the consequences of the heavy composition scenario for the search of UHECR sources. In particular, we show images predicted for some specific galaxy clusters and for the supergalactic plane. We find that heavy nuclei ``astronomy'' looks very different from what one expects for UHECR protons, with, for example, multiple images of the
same source, or unusual spatial distributions of events with different energies from a given source. This has important consequences for the reconstruction of UHECR source positions. Up to now, source reconstruction techniques often implicitly assume proton or light nuclei primaries, see, for example, Refs.~\cite{Golup:2009cv,Giacinti:2009fy}.

In Section~\ref{GMFmodels}, we review the regular GMF models we use in this study. In Section~\ref{Backtrace}, iron nuclei are backtraced in these GMF models. Distributions of angular densities of outgoing backtraced nuclei are computed in momentum space. In Section~\ref{AstrophysicalSources}, we focus on the images of astrophysical sources of heavy nuclei.

\section{Models of the regular Galactic Magnetic Field}
\label{GMFmodels}

\subsection{Models}
\label{Equations}

\begin{figure}
\begin{center}
\includegraphics[width=0.56\textwidth]{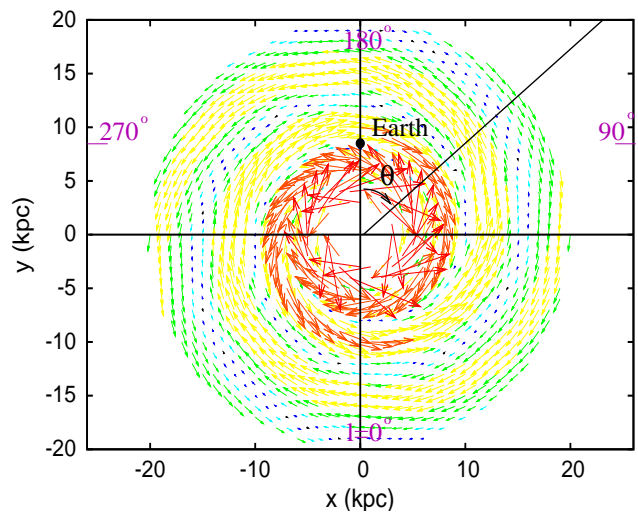}
\end{center}
\caption{Galactic plane ($z=0$) seen from $z>0$. Disk field of the PS model. For clarity, the field in the central region of the Galaxy is not displayed. Colors according to the strength of the field and arrows according to its direction.}
\label{PSangles}
\end{figure}

The PS model that we use contains three components: the disk, the toroidal field in the halo, and a central dipole.
We shall use Cartesian coordinates $x$, $y$ and $z$, Galactocentric cylindrical coordinates
$(r,\theta,z)$, where $r=\left(x^{2}+y^{2}\right)^{1/2}$, and Galactic coordinates $(l,b)$, respectively defined as in Fig.~\ref{PSangles}. The Earth is located at $(x=0,y=r_{\odot}=8.5\,{\rm kpc},z=0)$, with $\theta=0$ at the position of the Earth.

The field in the disk is parametrized in a way similar to the BSS-S version of the model of Ref.~\cite{Tinyakov:2001ir},
\begin{equation}
\begin{array}{ll}
B_{r} = B\left(r,\theta,z\right)\sin p,\\
B_{\theta} = B\left(r,\theta,z\right)\cos p\,,
\end{array}
\end{equation}
where we use as pitch angle $p=-8^{\circ}$. Furthermore,
\begin{equation}
B\left(r,\theta,z\right) = b\left(r\right) \cos\left[\theta-\frac{1}{\tan p} \ln\left( \frac{r}{\xi_{0}}\right) \right]\,\cdot\,\exp\left( -\frac{|z|}{z_{0}} \right)\,,
\end{equation}
where $z_{0}=0.2\,$kpc is the thickness of the thin disk, $b(r)$ is constant in the bulge and decays as $1/r$ outside,
\begin{displaymath}
b(r) \left\{ \begin{array}{ll}
                = \mbox{const}  & \mbox{for }r\leq 4\,{\rm kpc}\\
		\propto r^{-1} & \mbox{for }r > 4\,{\rm  kpc}
               \end{array} \right.\,,
\end{displaymath}
and
\begin{displaymath}
 \xi_{0}=\left(r_{\odot}+d\right) \exp\left( -\frac{\pi}{2} \, \tan p\right)\,,
\end{displaymath}
with $d=-0.5$\,kpc as the distance to the first field reversal.
The regular magnetic field strength is set to 2\,$\mu$G at the position of the Earth~\cite{Ringberg}.


The toroidal contribution ${\bf B}_T$ consists of two disks on each side of the galactic plane, with opposite signs, having their maximum strengths at $h_{T}=\pm1.5\,$kpc, Lorentzian widths of $w_{T}=0.3\,$kpc and exponentially decaying amplitudes beyond $r=r_{\odot}$,
\begin{equation}
\begin{array}{ll}
 B_{Tx} = -B_{T}~\mbox{sgn}\left(z\right)\cos\theta,\\
 B_{Ty} = B_{T}~\mbox{sgn}\left(z\right)\sin\theta\,,
\end{array}
\end{equation}
where
\begin{equation}
B_{T} = 1.5\,\mu\mbox{G} \, \frac{\Theta\left(r_{\odot} - r\right) + \Theta\left( r- r_{\odot} \right) \, \exp\left( \frac{r_{\odot} - r}{r_{\odot}}\right) }{1+\left( \frac{|z|-h_{T}}{w_{T}}\right)^{2}}\,,
\end{equation}
and $\Theta$ is the Heaviside step function.

The non-thermal filaments in the Galactic center support the existence of a poloidal contribution ${\bf B}_P$~\cite{Han:2009ts}. This third component can be modeled as in Ref.~\cite{Kachelriess:2005qm},
\begin{equation}
\begin{array}{ll}
B_{Px} = -3\frac{\mu_{D}}{R^3}\,\cos\phi~\sin\phi~\sin\theta\,,\\
B_{Py} = -3\frac{\mu_{D}}{R^3}\,\cos\phi~\sin\phi~\cos\theta\,,\\
B_{Pz} = \frac{\mu_{D}}{R^3}\,\left( 1-3\cos^{2}\phi\right)\,,
\end{array}
\label{PSdipole}
\end{equation}
where $R=\left(x^{2}+y^{2}+z^{2}\right)^{1/2}$, and $\cos\phi=z/R$. Observations suggest that at Earth the local vertical component of the magnetic field is $B_{z}\simeq0.2\,\mu$G~\cite{Han:2009jg}. If one assumes that it is entirely due to the dipole contribution, then $\mu_{D}=120\,\mu$G$\cdot$kpc$^{3}$~\cite{Kachelriess:2005qm}.

We also consider the case of a ten times weaker dipole, $\mu_{D}=12\,\mu$G$\cdot$kpc$^{3}$, or of a 
zero dipole, $\mu_{D}=0$, in the following sections.
The strength of the regular GMF in the Galactic center is not well constrained. Its order of magnitude is believed to be between tens of $\mu$G and possibly up to mG~\cite{Han:2009ts}. In order to avoid a singularity at the center, we cut each component of the field strength at a maximum value of 100\,$\mu$G, $|B_{x,y,z}|\leq100\,\mu$G. We verified that cutting $|B_{x,y,z}|$ at 10\,$\mu$G and at 1\,mG does not result in noticeable differences in the figures presented in the
following sections.


For comparison, we also present results of our simulations for two other GMF models.
For the first one we use the ``ASS+RING'' model presented in Ref.~\cite{Sun:2007mx} (called hereafter the ``Sun08 model'').
It is, among the different models these authors tested, the one which they found to be in best agreement with the data they considered.
In this model, the disk field is axisymmetric. The pitch angle is set to $p=-12^{\circ}$. Defining 
$B\left(r,\theta,z\right)=D_{1}\left(r,z\right)\,D_{2}\left(r\right)$, the functions are
\begin{equation}
D_{1}\left(r,z\right) = \left\{ \begin{array}{ll}
                B_{0}\exp\left( -\frac{r-r_{\odot}}{r_{0}}-\frac{|z|}{z_{0}}\right) & \mbox{for }r > r_{c}\\
		B_{c}\exp\left(-\frac{|z|}{z_{0}}\right)  & \mbox{for }r\leq r_{c}
               \end{array} \right.\,,
\end{equation}
and
\begin{equation}
D_{2}\left(z\right) = \left\{ \begin{array}{ll}
                +1 & \mbox{for }r > 7.5\,{\rm kpc}\\
		-1 & \mbox{for }6\,{\rm kpc} < r \leq 7.5\,{\rm kpc}\\
		+1 & \mbox{for }5\,{\rm kpc} < r \leq 6\,{\rm kpc}\\
		-1 & \mbox{for }r \leq 5\,{\rm kpc}
               \end{array} \right.\,,
\end{equation}
with
$B_{0}=B_{c}=2\,\mu$G, $r_{c}=5\,$kpc, $r_{0}=10\,$kpc and $z_{0}=1\,$kpc.
The toroidal contribution is a slightly modified version of the one in the PS model,
\begin{equation}
B_{T} = B_{T0} \cdot \frac{\frac{r}{r_{T0}} \exp\left( \frac{r_{T0} - r}{r_{T0}}\right) }{1+\left( \frac{|z|-h_{T}}{w_{T}}\right)^{2}}\,,
\end{equation}
with $B_{T0}=10\,\mu$G, $r_{T0}=4\,$kpc, $h_{T}=1.5\,$kpc, $w_{T}=w_{T,\rm in}=0.2\,$kpc for $|z|<h_{T}$ and $w_{T}=w_{T,\rm out}=0.4\,$kpc for $|z| \geq h_{T}$.

As explained in Ref.~\cite{Sun:2007mx}, the scale height of the thermal electron density may be underestimated by a factor of two, leading to such a high value of the halo field. Since the field in the halo is not well known, we decide to consider in this paper a third model, which is identical to the Sun08 model, but with different values for some of the halo parameters. This aims at showing the dependence of our results on the halo parameters. Reference~\cite{Sokoloff90}, which discusses 
the Galactic dynamo, suggests that the halo field may be much weaker, thicker and with a maximum strength further from the disk plane. We take here, as an example: $B_{T0}=1\,\mu$G, $h_{T}=4\,$kpc, $w_{T,in}=w_{T,out}=2\,$kpc. The value of $r_{T0}$ is left unchanged. Below, we refer to this model as the ``Sun08-MH (Modified Halo) model''.

No dipole contribution is added in the last two models. All three fields are set to zero for $r>20$\,kpc.

The detailed results on deflection maps are strongly model dependent. Nevertheless, in the following sections we are able to draw some general conclusions for the case of UHECR nuclei propagation in the GMF.


\subsection{Neglected effects}
\label{NeglectedContributions}

Since energy losses are negligible on Galactic scales, one usually backtraces nuclei
of opposite negative charge in the GMF at constant energy in order to map the arrival 
directions between the sky observed at Earth and the sky of arrival directions outside the Galaxy. 
Furthermore, we neglect any deflections in extragalactic magnetic fields (EGMF) in the present work:
Current simulations~\cite{EGMF} of the EGMF agree on the fact that these fields
tend to follow the large scale galaxy structure, i.e. the fields tend to be strongest
around the largest matter concentrations. However, they disagree on certain aspects that are
relevant for UHECR deflection, most notably the filling factor distributions, i.e.
the fraction of space filled with EGMF above a certain strength, as a function of
that strength~\cite{Sigl:2004gi}. While this causes considerable differences in the 
size of the deflection angles
predicted between the source and the observed events, the deflections tend to be
{\it along and within} the cosmic large scale structure of the galaxy distribution.
In other words, as long as the sources are not very nearby, the EGMF is unlikely
to deflect UHECRs out of the large scale structure since the fields in the voids
are very small. This means that the overall UHECR arrival direction distribution arriving outside the Galaxy is likely to still correlate with the local large scale structure even in
the scenarios with large EGMF, heavy nuclei and large deflection angles, although
the events do in
general not point back to the sources. On the other hand, since deflections in the
Galactic field are unlikely to correlate with extragalactic deflections, large deflections
of heavy nuclei in the Galactic field are expected to have a much stronger influence
on correlations with the local large scale structure. Within a first approximation
we can, therefore, restrict ourselves to the question how Galactic deflection changes
the images of extragalactic sources in the absence of EGMF, as long as we are
concerned with sources following the large scale structure as a whole.

\begin{figure}
\begin{center}
\includegraphics[width=0.49\textwidth]{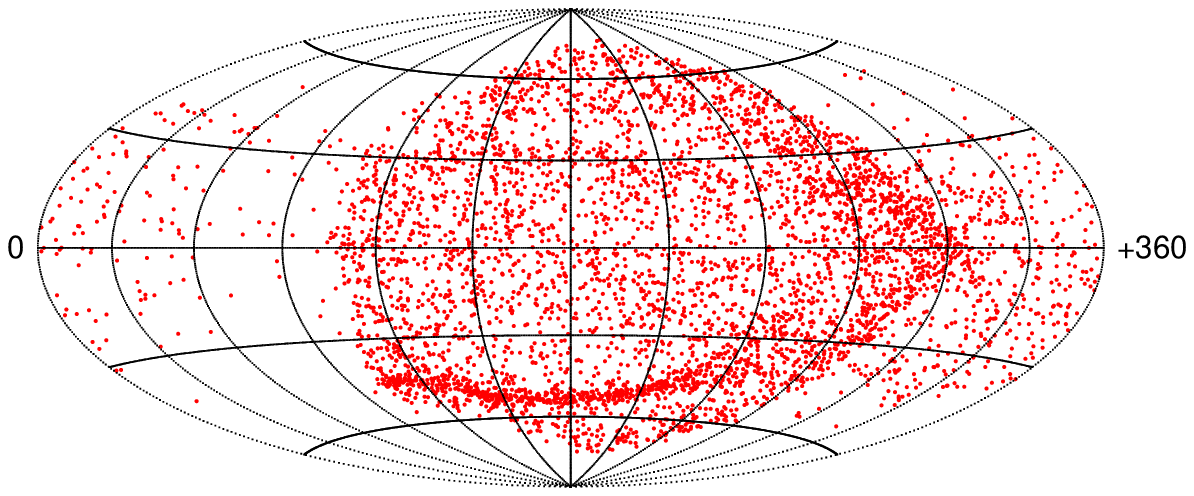}
\includegraphics[width=0.49\textwidth]{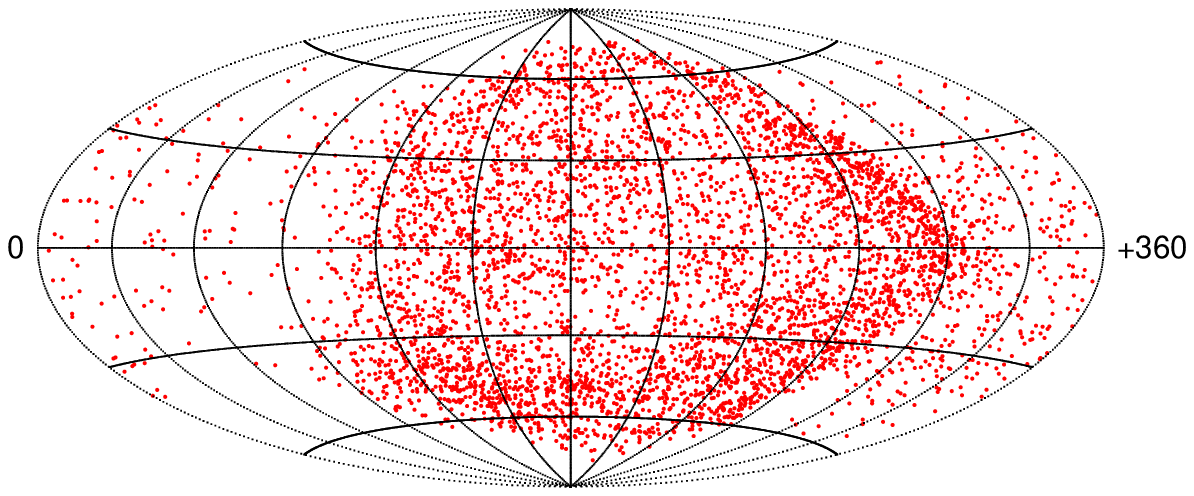}
\end{center}
\caption{\textbf{Left panel:} Final positions on the celestial sphere outside the Galaxy of 5000 iron nuclei of 60\,EeV injected isotropically around the Earth and backtraced in the PS model. Energies smoothed according to an experimental resolution $\Delta E/E=6$\%. \textbf{Right panel:} Same conditions, but with the turbulent Galactic field contribution added to the regular field. A field with a correlation length $L_{c}=50$\,pc and 1000 modes was assumed. The regular field has the dominant influence on the qualitative features of the backtraced distribution.}
\label{RegTurb}
\end{figure}

In Fig.~\ref{RegTurb}, we compare the angular positions of iron nuclei with energy $E=60$\,EeV 
backtraced from the Earth in the PS model of the regular GMF only (left panel) and in the same model combined
with the turbulent GMF (right panel). Figures~\ref{RegTurb} to \ref{MapDensOtherModels}, \ref{GC} and \ref{SGPPS} are in galactic coordinates.
The turbulent field was generated as a superposition of discrete wavenumber modes with
random polarization,
following the method presented in Ref.~\cite{Giacalone:1994}. For the profile of the
RMS amplitude of the field we used the toy model presented in Ref.~\cite{Giacinti:2009fy}. 
Taking a similar profile with an amplitude decaying as $1/r$, as for the regular PS GMF model, instead of the exponentially decaying profile in this toy model leads to qualitatively equivalent results. We used 1000 modes whose wavenumber all have the same length, corresponding to
a correlation length of $L_{c}=50$\,pc~\cite{Ringberg}.
Using a broader range of wavelengths with, for example, a Kolmogorov spectrum with lengths between 20\,pc and 200\,pc would not make noticeable changes on Fig.~\ref{RegTurb} (right panel).
While the turbulent field blurs the thin features present in the left panel of Fig.~\ref{RegTurb}, the general features of the angular images are not changed qualitatively by the turbulent component. Thus we conclude that it
is sufficient to consider only the regular component of the GMF for a qualitative discussion of 
propagation of nuclei in the GMF.
As we shall discuss in more detail later, one can already notice from Fig.~\ref{RegTurb} that some extragalactic regions are not reached by nuclei backtraced in the GMF considered here.
This implies that nuclei sources in these regions are not visible at Earth.

\section{Backtracing heavy nuclei in the Galactic magnetic field}
\label{Backtrace}

\begin{figure}
\begin{center}
\includegraphics[width=0.49\textwidth]{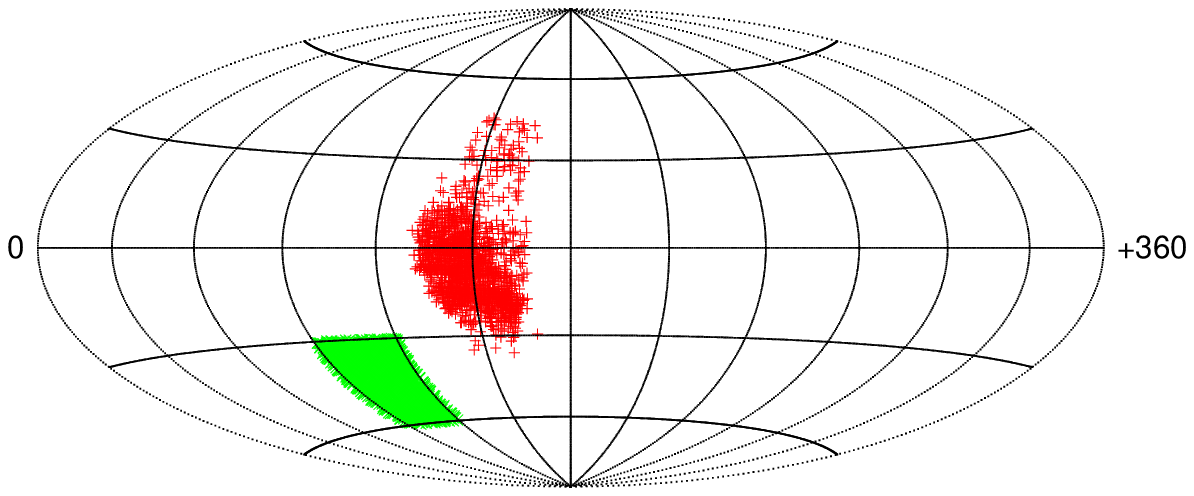}
\includegraphics[width=0.49\textwidth]{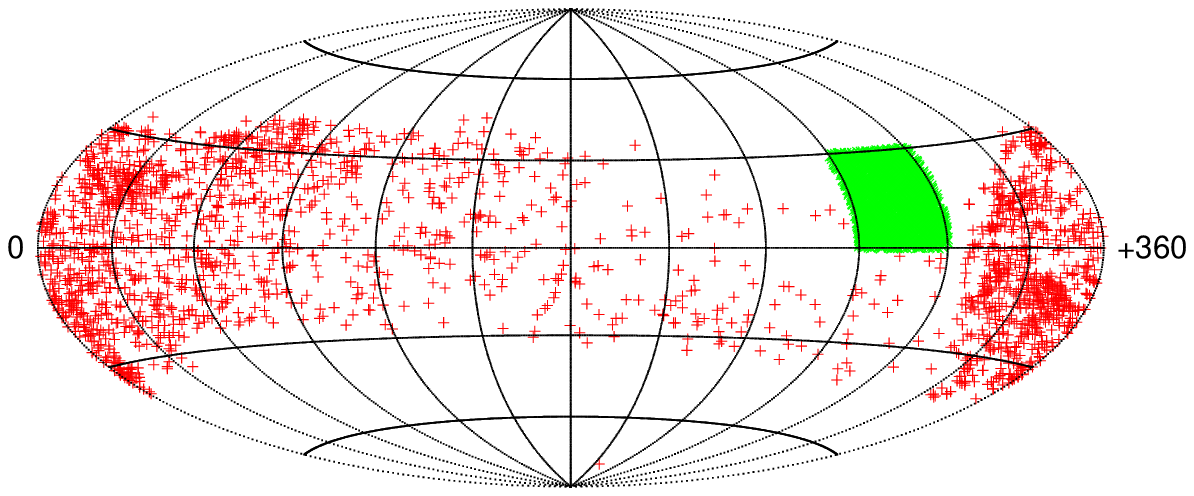}
\end{center}
\caption{Sky maps for 60\,EeV iron nuclei backtraced from the Earth to outside the Galaxy, in the PS model. The initial directions of the nuclei momenta at Earth are in green and the final momenta outside the Galaxy are in red.}
\label{indcellsPts}
\end{figure}

In this section, we backtrace iron nuclei in the three GMF models discussed in Section~\ref{GMFmodels} concentrating on the PS model. As noted in Fig.~\ref{RegTurb} (left panel) the backtracing of 60\,EeV iron nuclei with isotropically distributed arrival directions at the Earth leads to an anisotropic distribution of CRs entering the Galaxy. In particular, there exist empty regions, i.e.\ parts of the sky that do not contribute to the UHECR flux
on Earth for a given total number of observed (or simulated) UHECRs. Restricting the initial directions to 
smaller regions of the celestial sphere, one sees that some regions are shifted in a more or less coherent way 
as in Fig.~\ref{indcellsPts} (left panel), whereas other regions are spread over a large fraction of the sky, as 
shown in Fig.~\ref{indcellsPts} (right panel). The remarkable spread in Fig.~\ref{indcellsPts} (right panel) is mainly due to the poloidal component of the GMF. Hence, some iron nuclei reaching the Earth with directions within certain areas of the celestial sphere can either come from sources belonging to a well defined region of the sky, or can come from many different sources spread over a large fraction of the sky.

\begin{figure}
\begin{center}
\includegraphics[width=0.49\textwidth]{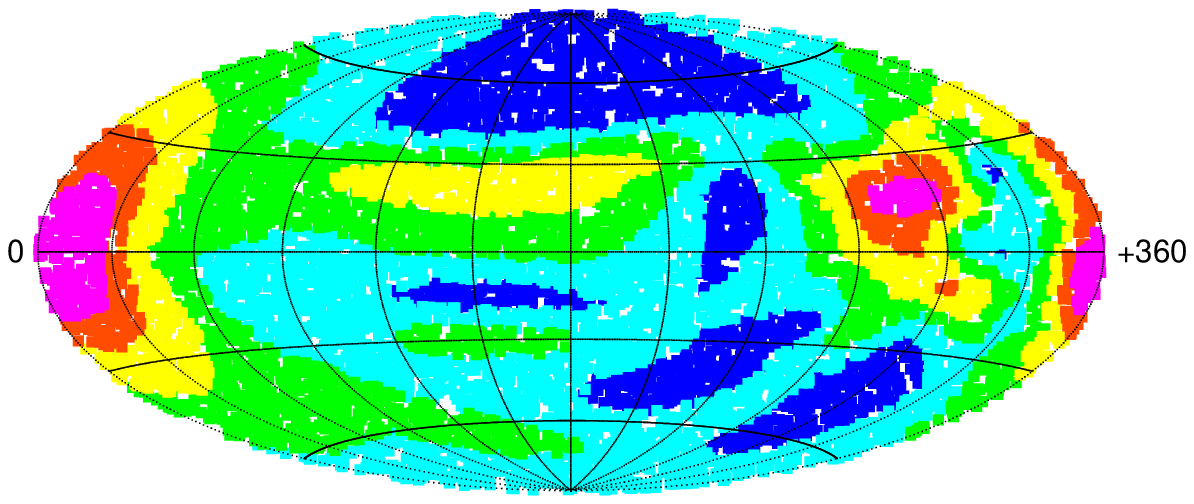}
\includegraphics[width=0.49\textwidth]{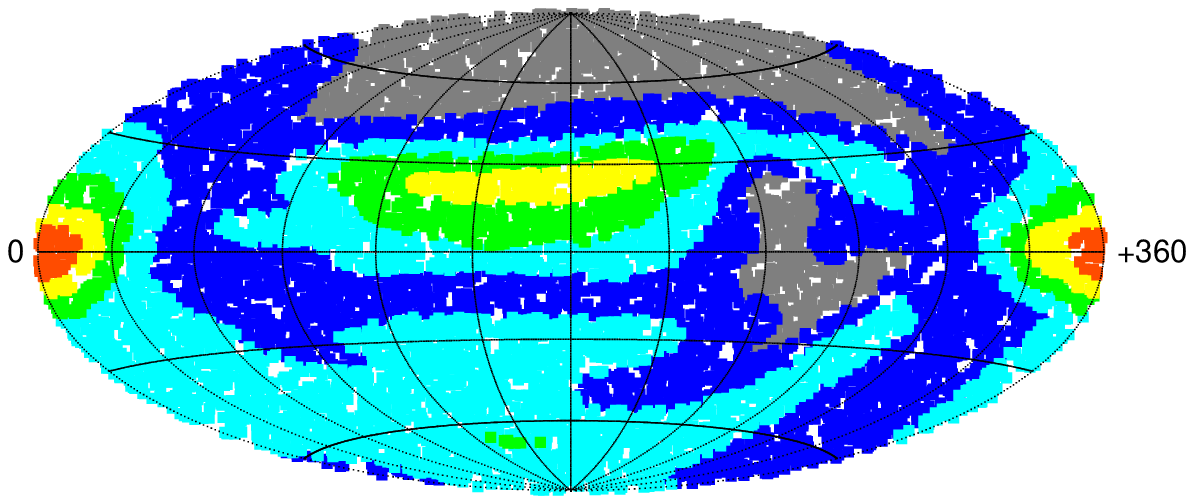}
\includegraphics[width=0.49\textwidth]{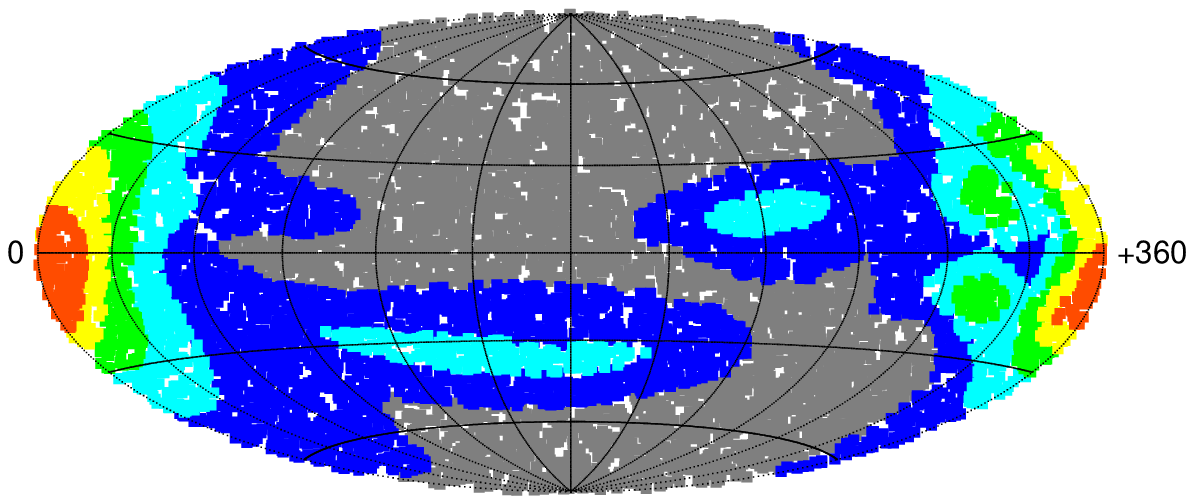}
\includegraphics[width=0.49\textwidth]{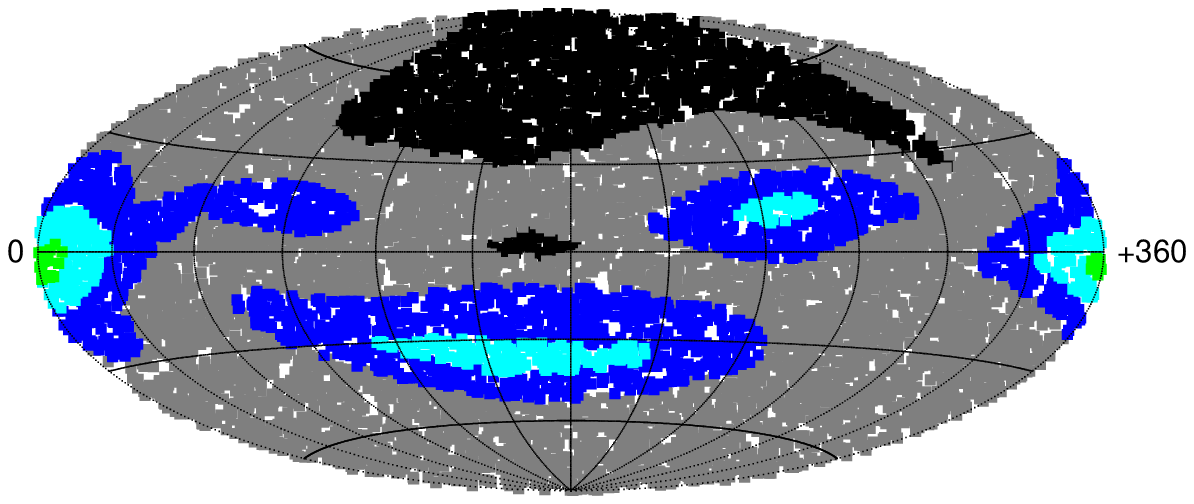}
\end{center}
\caption{Deflection angles in the PS model as a function of arrival direction detected at Earth. \textbf{Left column:} For $\mu_{D}=120\,\mu$G$\cdot$kpc$^{3}$. \textbf{Right column:} For $\mu_{D}=12\,\mu$G$\cdot$kpc$^{3}$. \textbf{Top row:} 60\,EeV iron. \textbf{Bottom row:} 140\,EeV iron. Black for angular deflections below 10$^{\circ}$, grey: 10$^{\circ}$-25$^{\circ}$, dark blue: 25$^{\circ}$-40$^{\circ}$, light blue: 40$^{\circ}$-55$^{\circ}$, green: 55$^{\circ}$-70$^{\circ}$, yellow: 70$^{\circ}$-85$^{\circ}$, orange: 85$^{\circ}$-100$^{\circ}$, magenta: $>$100$^{\circ}$.}
\label{Defl}
\end{figure}

We also map, for the PS model, the absolute values of deflection angles as a function of the arrival direction at Earth in Fig.~\ref{Defl}. These are angles, in degrees, between arrival directions outside the Galaxy and at Earth. We backtraced 10$^{5}$ iron nuclei of 60\,EeV (top row) and 140\,EeV (bottom row) in the PS model with $\mu_{D}=120\,\mu$G$\cdot$kpc$^{3}$ (left column) or with $\mu_{D}=12\,\mu$G$\cdot$kpc$^{3}$ (right column). We then computed on the sky map of arrival directions at Earth, the mean values of deflection angles for nuclei located within circles of 10 degrees radius covering the whole celestial sphere.
The chosen radius of $10^\circ$ is smaller than the expected mean deflections of ultrahigh energy nuclei in the GMF ($\sim50^{\circ}$ at 60\,EeV), but large enough to smooth small-scale features.
The deflections computed without dipole field, $\mu_{D}=0$, are very similar to those with 
$\mu_{D}=12\,\mu$G$\cdot$kpc$^{3}$.


The UHECR flux observed from a certain source is magnified or demagnified depending on the source position on the sky and the considered energy.
This effect can be important especially in case of UHECR heavy nuclei. 

There are two possible methods to compute the ``amplification'' factor $\mathcal{A}$. The first one is the triangulation method suggested in Ref.~\cite{Harari:1999it}. It consists in backtracing a triangle spanned by the momenta of three particles with nearby initial directions of momenta. The amplification factor $\mathcal{A}$ is then
the ratio of the initial area and its image at the border of the Galaxy. We have verified our numerical code by reproducing with this method the results for the PS model of Ref.~\cite{Kachelriess:2005qm}. For large deflections, this method becomes however impractical: The more distorted the image of the original triangle becomes, the more CRs have to be backtraced to obtain the correct border, and thus area, of the image. For sufficiently small rigidities, deflections become so strong that one enters the strong lensing regime. In this case, the topology of the backtraced area can change, and the triangulation method cannot give consistent results. In the second method, conceptually much simpler, one injects isotropically CRs,
backtraces them, and calculates their density outside the Galaxy. This method was used in Ref.~\cite{virgoMF} to study the spectrum and the anisotropy of UHECRs escaping from the
magnetic field of the Virgo cluster.
Since the flux at the Earth is taken to be isotropic, the backtraced densities outside the Galaxy
in units of the average density directly corresponds to the amplification factor $\mathcal{A}$.


Fig.~\ref{RegTurb} (left panel) shows that the angular density of backtraced points outside the Galaxy, when starting with a uniform density at Earth, strongly varies between different regions of the sky. For the plots presented in the following, we first backtrace between $5\times10^{4}$ and $4\times10^{5}$ iron nuclei. Then, on the celestial sphere of outgoing directions outside the Galaxy, we draw circles with radii $\mathcal{R}\leq20^\circ$ and compute the local angular density $\rho$ of backtraced nuclei within these circles.

Let us denote the mean angular density over the whole celestial sphere with $\left\langle\rho\right\rangle$. In the following figures and distributions, the values of $\log_{10}(\rho/\left\langle\rho\right\rangle)$ are shown. We also define a ``blind region'' as a strongly underdense region on the celestial sphere of outgoing directions at the edge of the Galaxy. We call regions ``blind'' when $\log_{10}(\rho/\left\langle\rho\right\rangle)<-2.5$, corresponding to $\rho/\left\langle\rho\right\rangle\la1/300$. A source placed in such a region would have its flux demagnified by more than a factor $\simeq300$. Neither the Pierre Auger Observatory nor future experiments like JEM-EUSO are expected to detect more than a few thousand events above 60\,EeV. This implies that sources in such regions are practically undetectable in the foreseeable future.

\begin{figure}
\begin{center}
\includegraphics[width=0.49\textwidth]{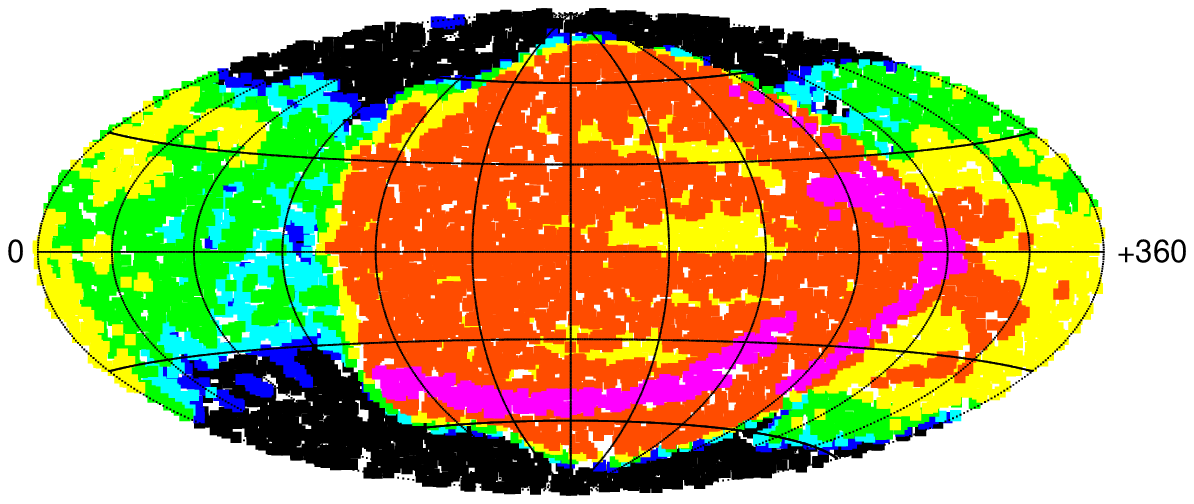}
\includegraphics[width=0.49\textwidth]{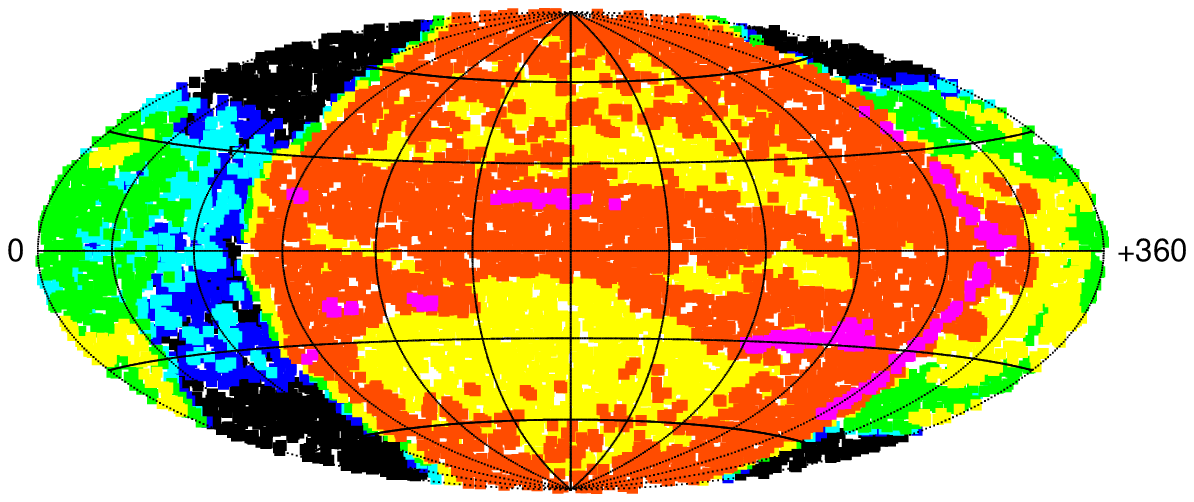}
\end{center}
\caption{Normalized densities outside the Galaxy of iron nuclei backtraced in the PS model, starting from an isotropic distribution at Earth. \textbf{Left panel:} $E=60\,$EeV. \textbf{Right panel:} $E=140\,$EeV. Colors represent the following density ranges: Dark blue for $-2<\log_{10}(\rho/\left\langle\rho\right\rangle)<-1.5$, light blue: for -1.5~to~-1, green: -1.~to~-0.5, yellow: -0.5~to~0, orange: 0~to~0.5, magenta: $>$0.5. Blind regions ($\log_{10}(\rho/\left\langle\rho\right\rangle)<-2.5$) in black.}
\label{MapDens}
\end{figure}

\begin{figure}
\begin{center}
\includegraphics[width=0.49\textwidth]{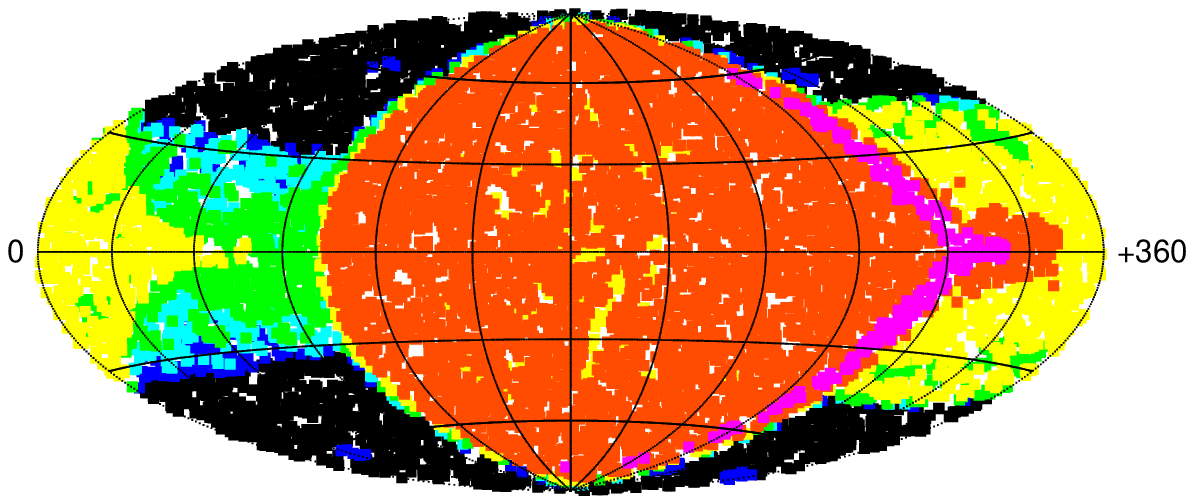}
\includegraphics[width=0.49\textwidth]{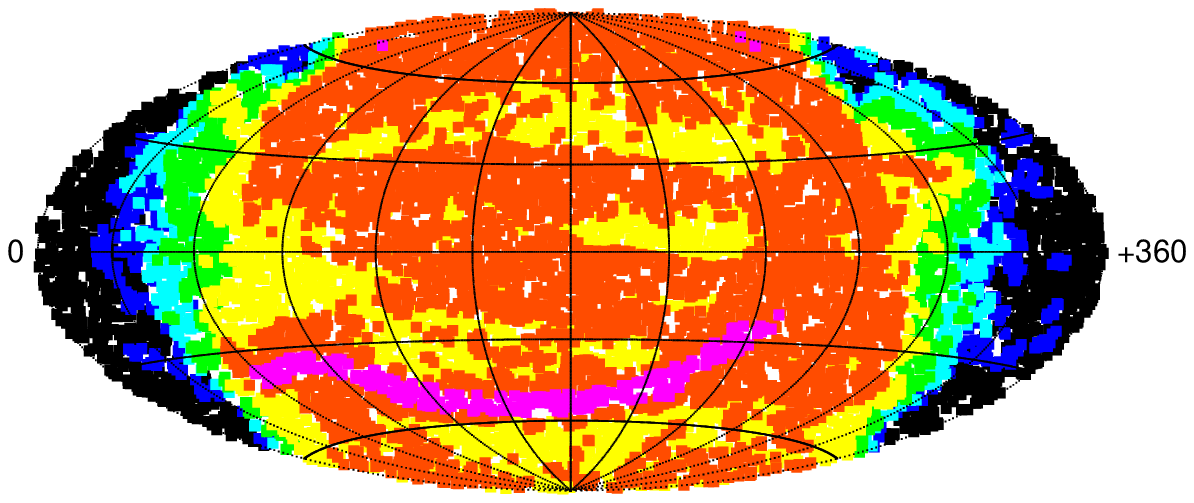}
\end{center}
\caption{Normalized densities of backtraced iron nuclei, outside the Galaxy: contributions of the PS model components at 60\,EeV. \textbf{Left panel:} Nuclei backtraced in the PS dipole only. \textbf{Right panel:} Nuclei backtraced in the PS model, without any dipole ($\mu_{D}=0$). Same key as in Fig.~\ref{MapDens}.}
\label{ContribDens}
\end{figure}

\begin{figure}
\begin{center}
\includegraphics[width=0.49\textwidth]{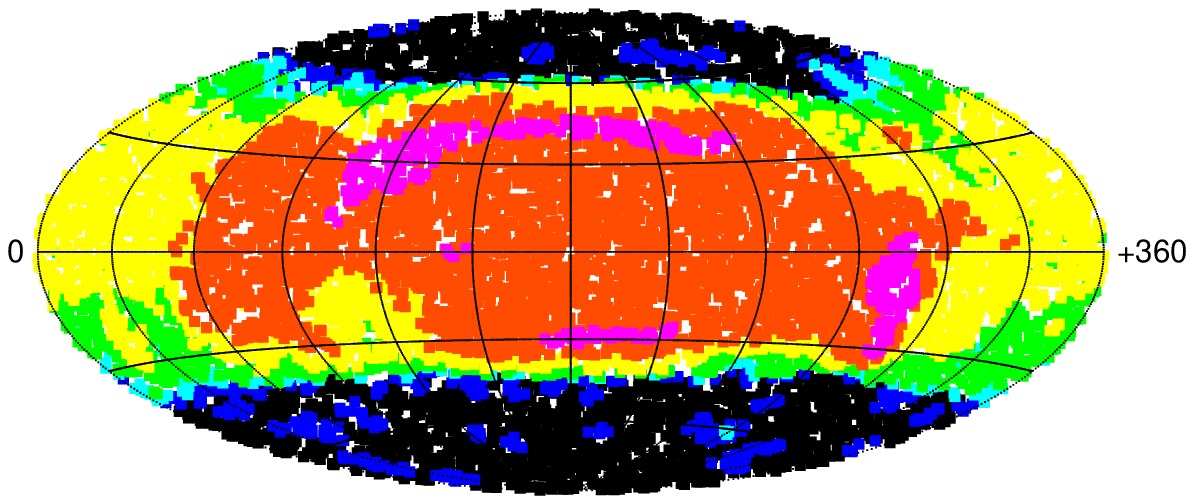}
\includegraphics[width=0.49\textwidth]{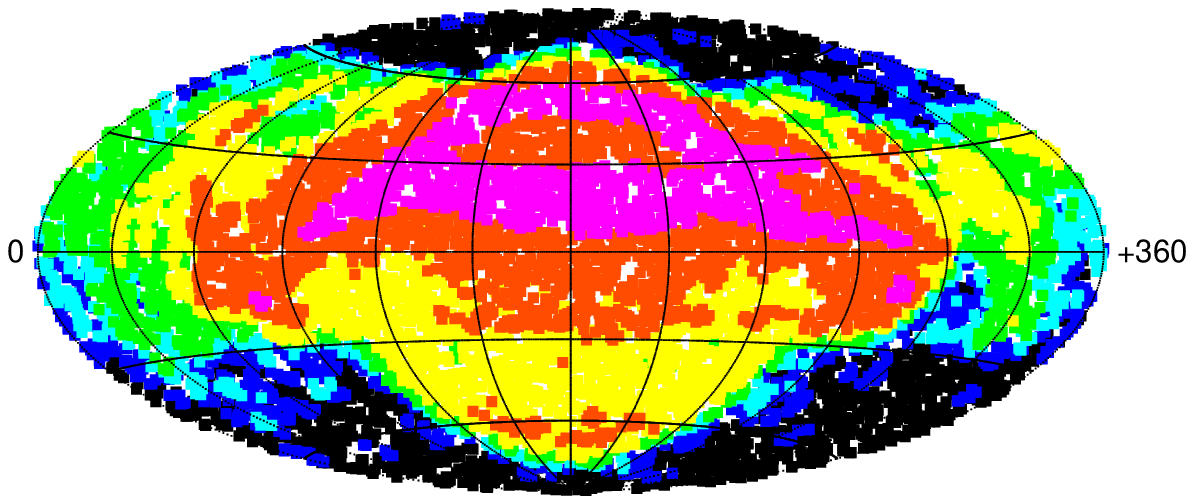}
\includegraphics[width=0.49\textwidth]{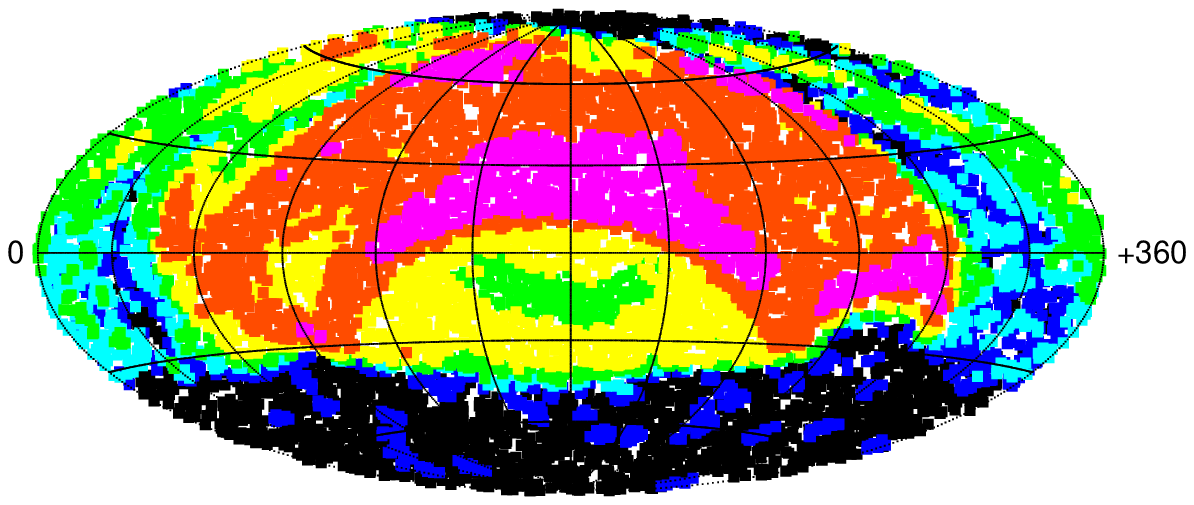}
\includegraphics[width=0.49\textwidth]{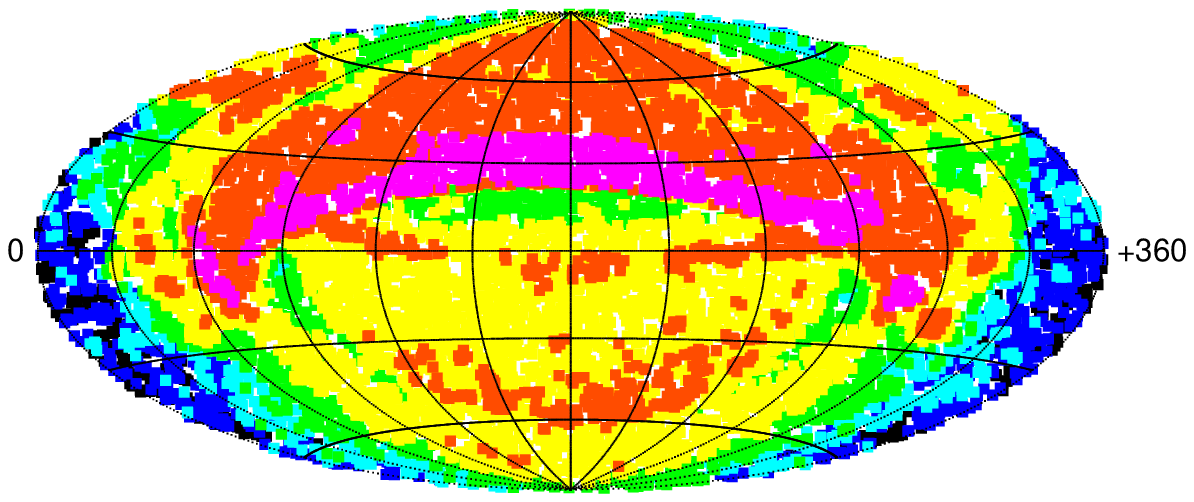}
\end{center}
\caption{Same as in Fig.~\ref{MapDens}, for the two other GMF models considered in this
paper. \textbf{Top row:} Sun08 model. \textbf{Bottom row:} Sun08-MH model. Same key as in Fig.~\ref{MapDens}.}
\label{MapDensOtherModels}
\end{figure}

In Figs.~\ref{MapDens}-\ref{MapDensOtherModels}, we plot sky maps of densities on the sphere of arrival directions outside the Galaxy for 10$^{5}$ backtraced iron nuclei. The densities are averaged over circles of $\mathcal{R}=3^\circ$ radius, and the whole sky is covered by 10000 of these circles.

Fig.~\ref{MapDens} presents sky maps of densities for backtraced iron nuclei in the PS model, outside the Galaxy, starting from an isotropic distribution at Earth. The energy of the nuclei
are 60\,EeV (left panel) and 140\,EeV (right panel), respectively.
10$^{5}$ iron nuclei were backtraced to produce these figures. Colors show different density bins: Dark blue stands for $-2<\log_{10}(\rho/\left\langle\rho\right\rangle)<-1.5$, light blue for $-1.5$~to~$-1$, green for $-1$~to~$-0.5$, yellow for $-0.5$~to~$0$, orange for $0$~to~$0.5$ and magenta for $>0.5$. Blind regions ($\log_{10}(\rho/\left\langle\rho\right\rangle)<-$2.5) are in black. First, one can notice that there is a big fraction of the sky for which densities do not differ by more than a factor three from the average density. Second, there is a significant fraction of the sky from which nuclei practically cannot reach the Earth.

In Fig.~\ref{ContribDens}, we plot the densities for iron nuclei backtraced in the dipole described in Eq.~(\ref{PSdipole}) only, and for nuclei backtraced in the PS model without the dipole, $\mu_{D}=0$. Comparing with Fig.~\ref{MapDens} (left panel), one can see that globally deflections in the PS model are dominated by the dipole.

Figure~\ref{MapDensOtherModels} shows the results corresponding to Fig.~\ref{MapDens} for the Sun08 and Sun08-MH models. The comparison of these maps for different models demonstrates that densities in any given direction of the sky are very model dependent.

\begin{figure}
\begin{center}
\includegraphics[width=0.49\textwidth]{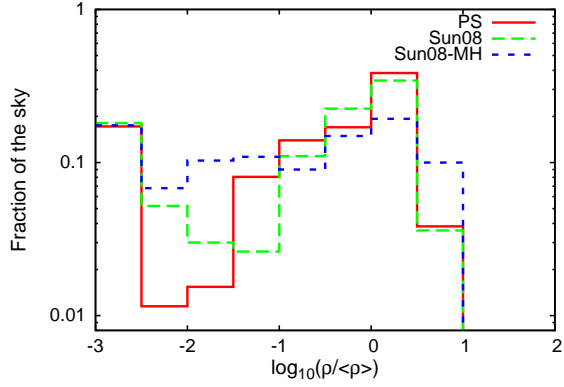}
\end{center}
\caption{Histogram of fractions of the sky outside the Galaxy with event densities within bins of width $\Delta\log_{10}(\rho/\left\langle\rho\right\rangle)=0.5$, for $4\times10^{5}$ backtraced 60\,EeV iron nuclei in the PS, Sun08 and Sun08-MH models, with $\Delta E/E=6\%$. Solid red line for the PS model, green dashed and blue dotted lines respectively for the Sun08 and Sun08-MH models. The bin of densities below $10^{-2.5}\left\langle\rho\right\rangle$ corresponds to blind regions.}
\label{Density}
\end{figure}

Figure~\ref{Density} gives the fractions of the sky with log$_{10}(\rho/\left\langle\rho\right\rangle)$ within a given interval, for the map of arrival directions outside the Galaxy and for $4\times10^{5}$ backtraced 60\,EeV iron nuclei with $\Delta E/E=6\%$.
The density distribution is not symmetric, with a larger tail for low densities than for high densities. The most populated density bin corresponds to a slight overdensity, with $0<\log_{10}(\rho/\left\langle\rho\right\rangle)<0.5$. In the first bin, $\log_{10}(\rho/\left\langle\rho\right\rangle)<-2.5$, all contributions with $\rho<10^{-2.5}\left\langle\rho\right\rangle$ are included. Thus the first bin gives the fraction of blind regions.

This distribution is obtained by computing the considered fraction of the sky from the number of circles of a given radius $\mathcal{R}$ which contain an event density falling within the given density bin. For densities close to the mean density, the exact value of the radius does not significantly affect the results in the range $2^\circ\leq\mathcal{R}\leq10^\circ$. The values of fractions of the sky in each density bin are fitted with a second order polynomial function, $a_{1}\mathcal{R}^{2}+a_{2}\mathcal{R}+a_{3}$, which gives an extrapolation of the fraction of the sky to the limit $\mathcal{R} \rightarrow 0 $. These extrapolated values are plotted in Fig.~\ref{Density}. The uncertainties are assumed to scale as $1/\sqrt{N}$, with $N$ being the number of events in the considered individual circles.

\begin{figure}
\begin{center}
\includegraphics[width=0.49\textwidth]{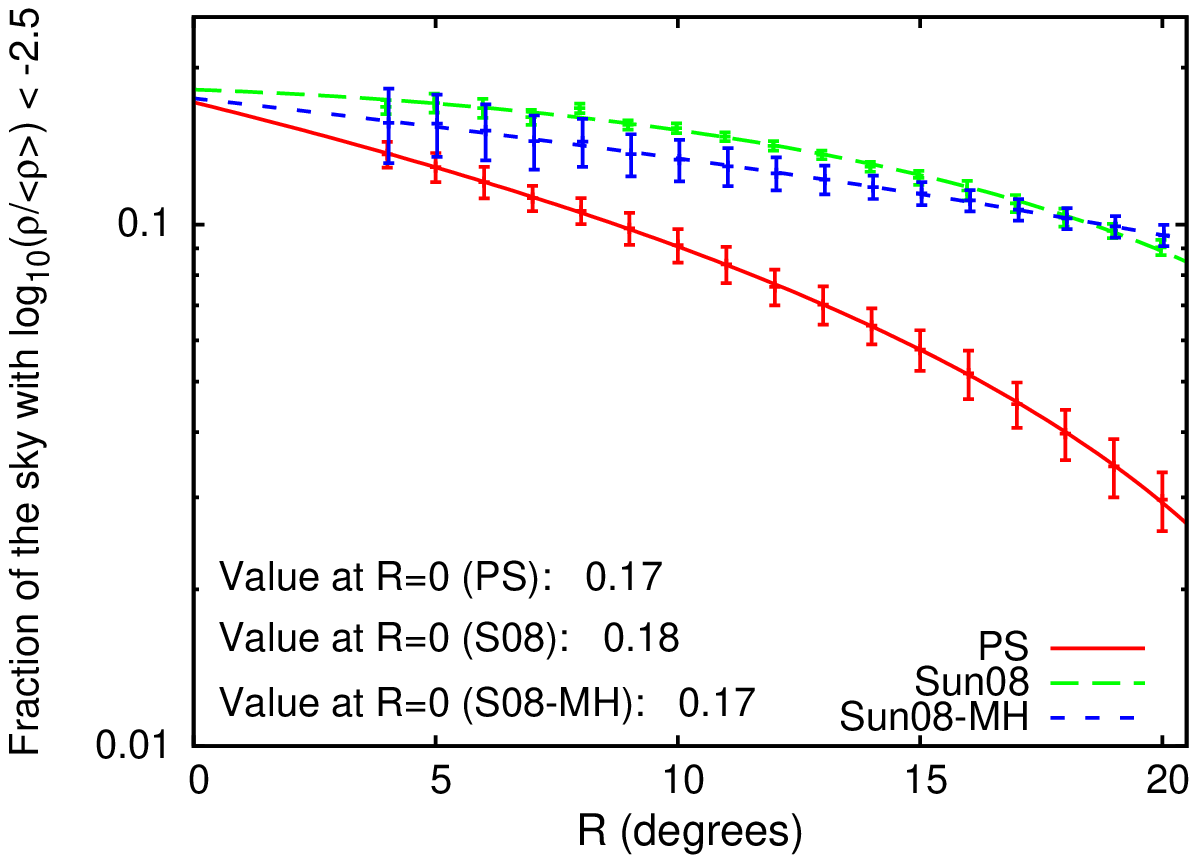}
\includegraphics[width=0.49\textwidth]{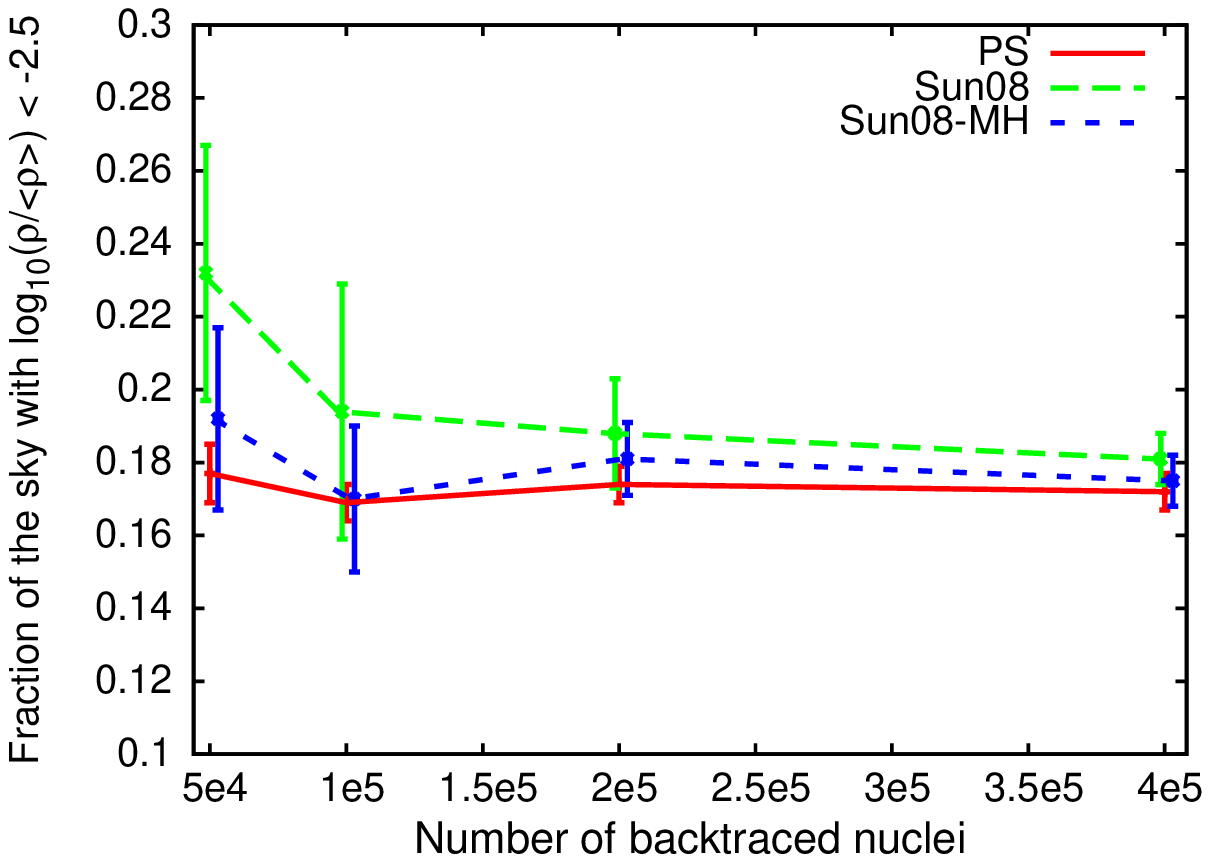}
\end{center}
\caption{\textbf{Left panel:} Blind fraction of the sky for the three models with $4\times10^{5}$ nuclei, versus the radii of the 'test' circles. Extrapolation of the real blind fraction of the sky, corresponding to zero degree radii circles. Solid red line for the PS model, green dashed and blue dotted lines respectively for the Sun08 and Sun08-MH models. \textbf{Right~panel:} Fraction of the sky outside of the Galaxy which is blind versus the number of 60\,EeV iron nuclei backtraced from the Earth. The energy resolution $\Delta E/E$ is set to 6\%.}
\label{BlindRegions}
\end{figure}

Since we decided to define blind regions as regions with a relative density smaller than a finite threshold
value, their size on the sky does not depend on the number of backtraced nuclei. However, as shown in Fig.~\ref{BlindRegions} (right panel), for too small numbers of nuclei on the sky (roughly below 2$\times10^{5}$), the method used here starts to give less precise results due to larger fluctuations.

The values plotted in Fig.~\ref{BlindRegions} (right panel) correspond the fraction of the sky with $\log_{10}(\rho/\left\langle\rho\right\rangle)<-2.5$, after interpolating as above to zero radius. The left panel shows such fits for the three considered models, and for $4\times10^{5}$ nuclei.


The fractions of blind regions for 60\,EeV iron nuclei are all close to 20\% for the three models we consider. These blind regions play an important role because, in a given energy range, the sources of heavy nuclei located in such regions do not contribute to the flux of UHECR collected at Earth. For example, in the PS model, the Virgo and Coma galaxy clusters cannot be seen at Earth for energies of 60-70\,EeV.

\section{Search for astrophysical sources of heavy nuclei}
\label{AstrophysicalSources}

The deflection angles of nuclei in the GMF are expected to be large, even at the highest energies, see Fig.~\ref{Defl}. For that reason, one of the most challenging questions in the case of  heavy nuclei as UHECRs is how to reconstruct the location of their sources. Because of photo-disintegration, the sources must be located in the local Universe. In the present section we construct images of nearby galaxy clusters, and of the whole supergalactic plane, where most of the potential UHECR sources should be located.

The general features of images of proton sources are well known and have already been studied in the context of source searches, see Refs.~\cite{Golup:2009cv,Giacinti:2009fy} (except in the unfavorable condition of sources in the Galactic plane). One could naively expect that images of heavy nuclei sources would display the same features, with angular scales enlarged by a factor $Z$. However, this simple case is rare, and features more complicated than for $Z=1$ often appear.


We consider here nearby galaxy clusters as effective extended bright sources of heavy nuclei. We do not consider the internal distributions of sources, but rather sum up all their individual contributions. The clusters are assumed to have a disk-like shape on the celestial sphere, with a typical radius of 5~degrees. We backtraced iron nuclei with different energies ranging from 60~EeV to 140~EeV, and recorded the directions of those which escape the Galaxy in a direction within an angular distance to the considered cluster of 5 degrees.

\begin{figure}
\begin{center}
\includegraphics[width=0.49\textwidth]{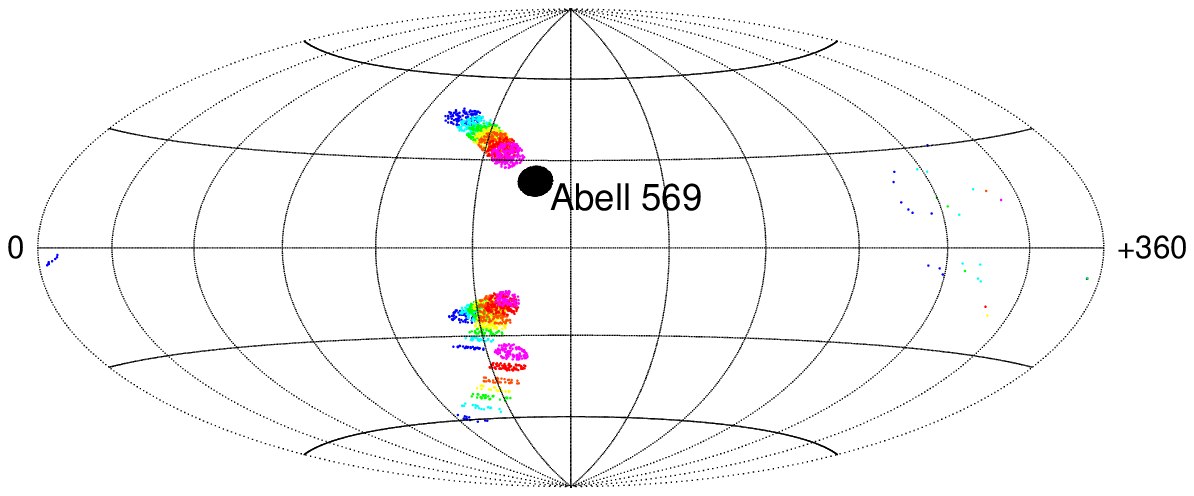}
\includegraphics[width=0.49\textwidth]{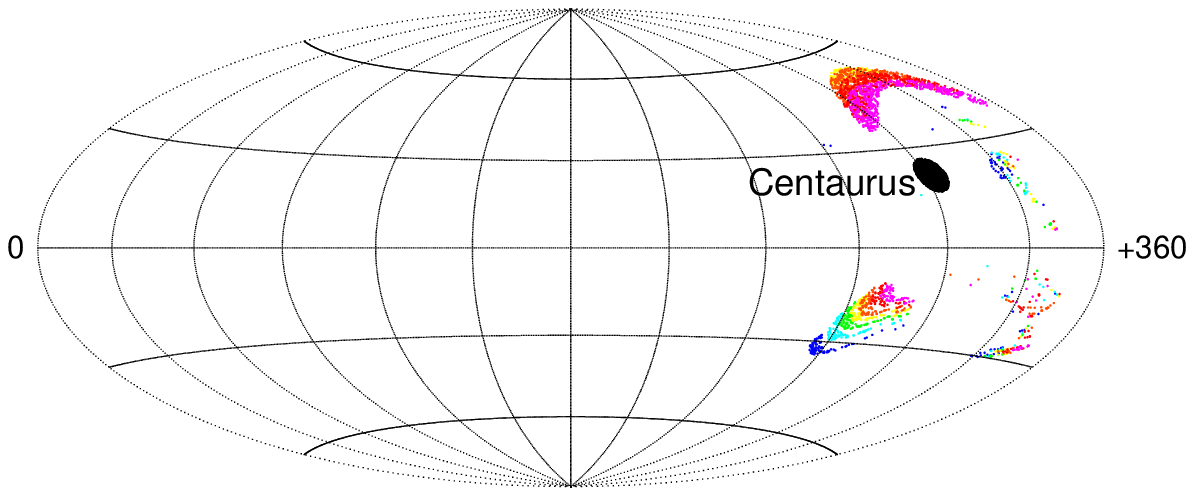}
\includegraphics[width=0.49\textwidth]{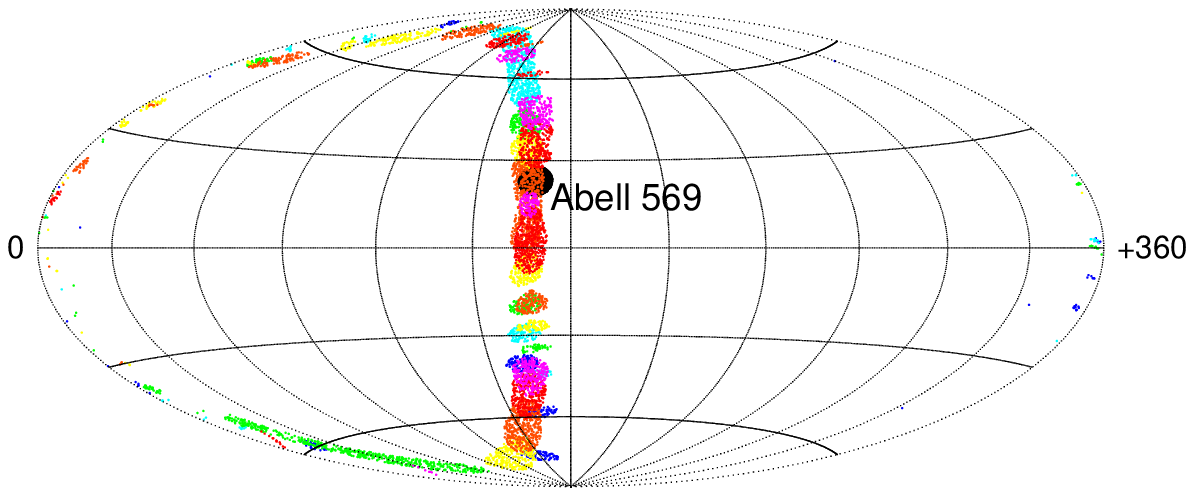}
\includegraphics[width=0.49\textwidth]{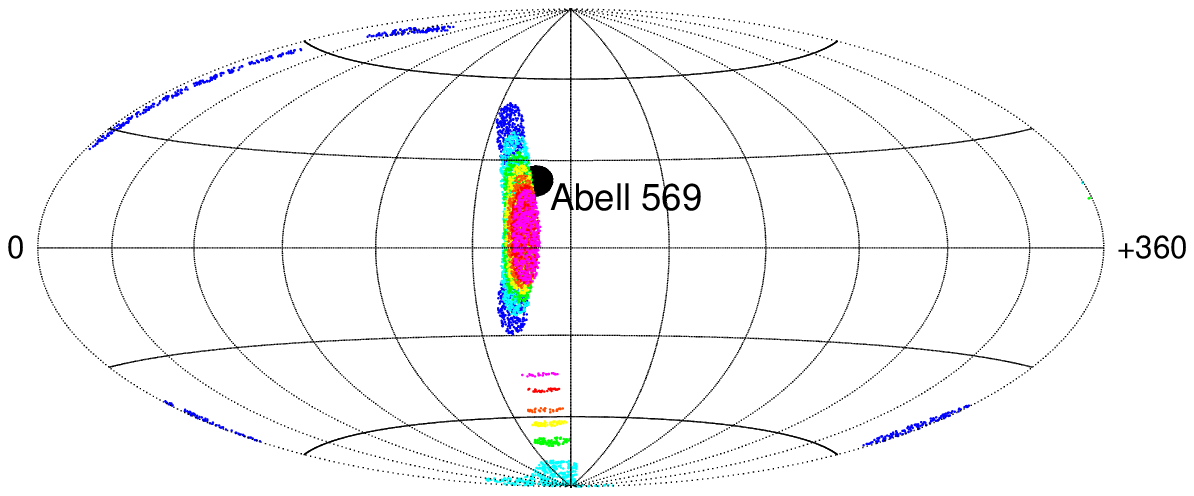}
\end{center}
\caption{Images of two nearby galaxy clusters emitting high energy iron (from 60\,EeV to 140\,EeV) deflected in GMF models. \textbf{Upper left panel:} Abell~569 in the PS model. \textbf{Upper right panel:} Centaurus in the PS model. \textbf{Lower left panel:} Abell~569 in the Sun08 model. \textbf{Lower right panel:} Abell~569 in the Sun08-MH model. Dark blue for 60\,EeV, light blue: 70\,EeV, green: 80\,EeV, yellow: 90\,EeV, orange: 100\,EeV, red: 120\,EeV, magenta: 140\,EeV.}
\label{GC}
\end{figure}

In Fig.~\ref{GC} (top row) we present the images at different energies of two nearby clusters, Abell~569 and Centaurus, for the PS model, and in Fig.~\ref{GC} (bottom row) the images of Abell~569 for the two other GMF models considered by us. The images of individual clusters can have very peculiar features even at the highest energies. For instance, they can have several images, similar to the case of strong gravitational lensing. This can be explained by the fact that when one backtraces a regular grid of iron nuclei from the Earth to outside the Galaxy, the grid outside the Galaxy is folded in several regions~\cite{Harari:1999it}. Then, one direction outside of the Galaxy can correspond to several directions at Earth, and one source can have multiple images. In practice, the nuclei corresponding to different images at Earth of the same source enter the Galaxy in different positions of the physical space, but with the same angles. In the cases presented in Fig.~\ref{GC}, the two galaxy clusters have several images.

Sometimes several images can merge into one single image when the energy is increased, as in Fig.~\ref{GC} (upper left panel), or an image can only appear above or below a certain threshold energy, for example, as the image at high energy and high latitudes of Centaurus in Fig.~\ref{GC} (upper right panel). In the PS model, some clusters like Coma and Virgo start to have images only above a given threshold energy, because they are located in a blind region for lower energies.

We also find that the deflections of the UHECR within a given image are usually distributed in a more complicated manner than the approximate $1/E$ behavior close to the ballistic regime.
In some particular cases, lower energy events can even be closer to the source than the highest energy events, which can be challenging for source reconstruction algorithms.

The images can also be strongly distorted. The upper panels of Fig.~\ref{GC} display distorted images of Abell~569 and Centaurus which are spread over Galactic longitudes between $\sim280^{\circ}$ and $\sim360^{\circ}$ because of the dipole. Sources for which no localized image exists can hardly be detected.

The two lower panels of Fig.~\ref{GC} show the images of Abell~569 in the Sun08 and Sun08-MH model. The localization and the shapes of the images of heavy nuclei sources are model dependent. Since the GMF is poorly known, this stresses the fact that none of these examples should be regarded as a prediction. With iron, one first needs a better knowledge of the GMF than is currently available, including in particular a better knowledge of the halo field. Among other factors, just considering different extensions and strengths of the halo field can already strongly change the general shape of source images, as shown in the two lower panels of Fig.~\ref{GC}. 
Extragalactic magnetic fields may further modify the images of individual sources.


\begin{figure}
\begin{center}
\includegraphics[width=0.49\textwidth]{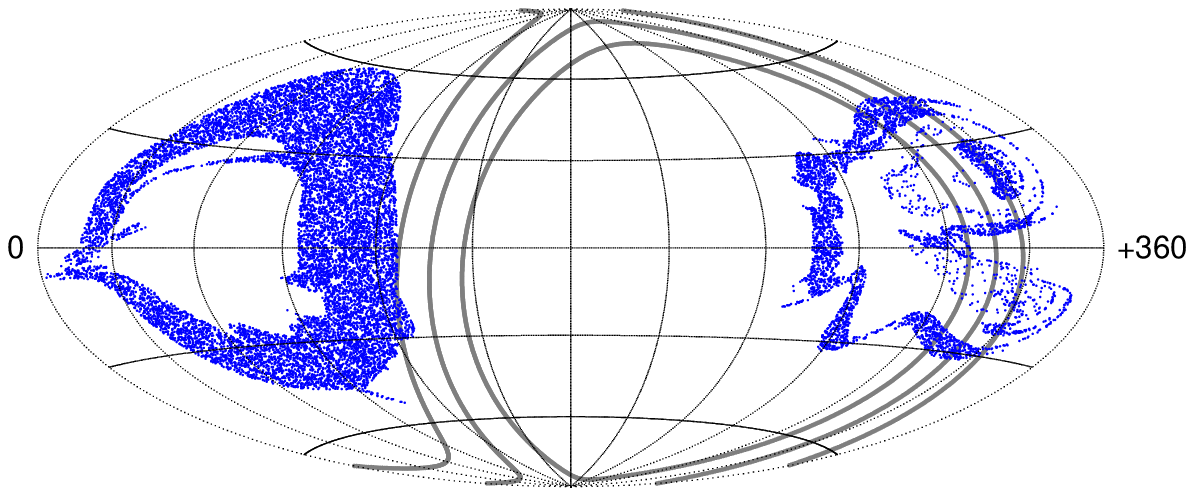}
\includegraphics[width=0.49\textwidth]{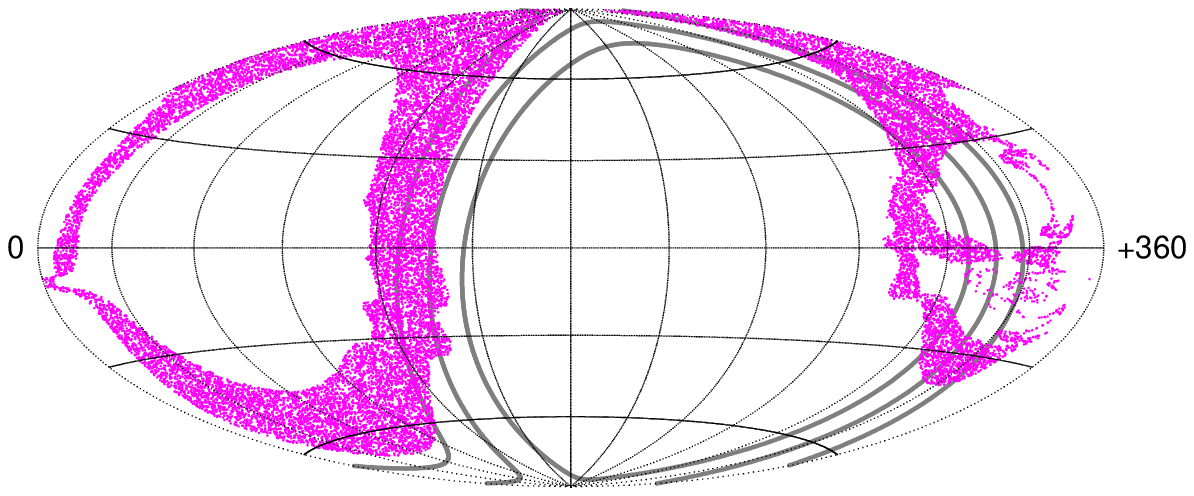}
\end{center}
\caption{Images of the supergalactic plane in the PS model. Thickness of the plane taken as $\pm10^\circ$ in supergalactic latitude. \textbf{Left panel:} In 60\,EeV iron nuclei. \textbf{Right panel:} In 140\,EeV iron nuclei. Grey central line for the supergalactic plane (sgb$=0^{\circ}$), between its two delimiting external grey lines (sgb$=\pm10^{\circ}$).}
\label{SGPPS}
\end{figure}

For the supergalactic plane, we assumed as in Ref.~\cite{Stanev:2008sd} that the plane region containing most of the local distribution of matter, and thus most of the potential astrophysical UHECR sources, has supergalactic latitudes between $\pm 10$ degrees. Most of the galaxy clusters considered above are of course located in or close to the supergalactic plane. To obtain images of the supergalactic plane, we proceed in a similar fashion as for galaxy clusters.

The images of the supergalactic plane are plotted in Fig.~\ref{SGPPS}, with 60\,EeV and 140\,EeV iron nuclei propagated in the PS model. It displays in fact two disconnected images at 60\,EeV. They are separated due to the dipole contribution: Without the dipole, the two images would merge into one.

In this scenario, at 60\,EeV, the parts of the plane which are located close to the Galactic polar regions cross ``blind regions'', see Section~\ref{Backtrace}. The sources located in these parts do not contribute to the UHECR flux reaching the Earth. This virtually ``cuts'' the supergalactic plane into two parts. In Fig.~\ref{SGPPS} (left panel), the ``left'' part of the supergalactic plane is responsible for the ``left'' image, and the ``right'' part for most of the ``right'' image. However, a very small fraction of the ``right'' image is produced by UHECR heavy nuclei produced in the ``left'' part of the supergalactic plane: in this model, this region of the sky is highly distorted, as already shown in Fig.~\ref{indcellsPts} (right panel).

The parts of the image which are close to the supergalactic plane do not necessarily involve small deflections, since some of them are considerably deflected {\it within\/} the supergalactic plane. 
Similarly, deflections in extragalactic magnetic
fields are unlikely to modify the image of the supergalactic plane as a whole since
such fields would mostly lead to deflections {\it inside\/} the large scale structure.

One can conclude, that in general no small angle correlation method which assumes that the arrival directions of cosmic rays are close to the position of their sources can be used to study nuclei sources.

\section{Conclusions and Perspectives}
\label{Conclusions}

We have investigated the propagation of heavy nuclei with energies above 60\,EeV in regular Galactic Magnetic Field models, especially in the Prouza and Smida model~\cite{Kachelriess:2005qm,PS}. The turbulent magnetic field contribution has been neglected as a first approximation, since its impact would not strongly change the qualitative results presented here. We also neglected deflection in extragalactic magnetic fields.

In Section~\ref{Backtrace}, we have backtraced in GMF models iron nuclei isotropically distributed around the Earth. We have shown that the sky maps of directions of backtraced nuclei momenta, outside the Galaxy, contain a wide range of angular densities: Extragalactic sources contribute very non-uniformly to the (isotropic) UHECR flux 
on Earth, with under- and overdense regions corresponding to (de-) magnification of the UHECR flux from sources located in such parts of the sky. In particular, there are also large regions of the sky which can be considered empty for a given number of backtraced nuclei: Regions with an underdensity $\log_{10}(\rho/\left\langle\rho\right\rangle)<-2.5$ can encompass one fifth of the sky, but would contribute only few events to the UHECR flux reaching the Earth even in a high-statistics with few thousands of events. We have noted also that some regions can be highly distorted, especially towards the Galactic center in the PS model.

We have computed the images at different energies of individual nearby galaxy clusters, assuming iron nuclei primaries arriving outside the Galaxy. We have also calculated the image of the supergalactic plane, where most of the astrophysical sources of UHECR should be located. 
In case of iron nuclei as UHECR primaries, the supergalactic plane
has essentially no overlap with its image at energies $E=60\,$EeV, and even 
at energies $E=140\,$EeV the images of a large fraction of CRs from sources in the supergalactic plane are shifted outside this plane.

The detailed deflection maps are very model dependent and should not be regarded as predictions. However, the distributions of fractions of the sky outside the Galaxy from which
the arriving UHECR flux is amplified or de-amplified by a certain amount is considerably less
model dependent. These results, therefore, provide some general ideas about the challenges of UHECR ``astronomy'' with heavy nuclei if such a heavy composition is confirmed.

The strong dependence of deflection maps on the GMF models, together with the fact that no current model fits correctly all observational data, leads to the conclusion that a better knowledge of this field is crucial for the search and identification of heavy nuclei sources. In particular, a more precise knowledge of the halo field strength, its extension and polarization, is required in case of a heavy UHECR composition.

Future radio telescopes, such as SKA and its precursor LOFAR, will significantly improve our knowledge on the structures and strengths of the Galactic and extragalactic magnetic fields. They will also provide new insights into their origin and evolution.
For example, SKA will provide an all-sky survey of rotation measures (RM) and increase by a few orders of magnitude the number of RMs, which are currently sparse, especially outside the Galactic plane. After one year of observation, one expects measures for approximately $2\times10^{4}$ pulsars and $2\times10^{7}$ compact polarized extragalactic sources. This will allow one to map out the global geometry of the GMF both in the halo and the disk, as well as its turbulent properties~\cite{Gaensler:2004gk,Beck:2004gq}.

\section*{Acknowledgments}
This work was supported by the Deutsche Forschungsgemeinschaft through the collaborative research centre SFB 676 Particles, Strings and the Early Universe: The Structure of Matter and Space-Time and by the State of Hamburg, through the Collaborative Research program Connecting Particles with the Cosmos within the framework of the Landesexzellenzinitiative (LEXI).

\end{document}